\documentclass[pra,twocolumn,showpacs]{revtex4-1}

\usepackage[utf8x]{inputenc}
\usepackage{amsmath,amssymb,amsfonts,subfigure,braket,color,dsfont}
\usepackage[english]{babel}
\usepackage[dvips]{graphicx}
\usepackage{sistyle}

\usepackage{subfig}

\renewcommand{\vec}[1]{\boldsymbol{#1}}

\newcommand{\M}{\mathcal{M}}

\DeclareMathOperator{\Expectation}{\mathbb{E}}

\DeclareMathOperator{\trace}{Tr}

\newcommand{\tensorp}{\otimes}
\newcommand{\id}{\mathds{1}}

\newcommand{\cm}[1]{\left[ #1 \right]}
\newcommand{\acm}[1]{\left\{ #1 \right\}}

\newcommand{\info}[1]{\textcolor{blue}{(#1)}}

\renewcommand{\info}[1]{}
\usepackage[draft]{fixme}

\begin{document}
\title{Bayesian parameter inference from continuously monitored quantum systems}

\author{S{\o}ren Gammelmark}
\author{Klaus M{\o}lmer}
\affiliation{Lundbeck Foundation Theoretical Center for
Quantum System Research, Department of Physics and Astronomy,
University of Aarhus, DK 8000 Aarhus C, Denmark}

\begin{abstract}
We review the introduction of likelihood functions and Fisher information in classical estimation theory, and we show how they can be defined in a very similar manner within quantum measurement theory.
We show that the stochastic master equations describing the dynamics of a quantum system subject to a definite set of measurements provides likelihood functions for unknown parameters in the system dynamics, and we show that the estimation error, given by the Fisher information, can be identified by stochastic master equation simulations.
For large parameter spaces we describe and illustrate the efficient use of Markov Chain Monte Carlo sampling of the likelihood function.
\end{abstract}

\pacs{02.50.Tt, 03.65.Yz, 42.50.Lc}
%
\maketitle

\section{Introduction}

Sensors and measurement devices are affected by the presence or strength of physical effects that influence their dynamics in a detectable way.
A proper statistical treatment of measurement data is important when inferring results from complex experiments.
With the growing use of quantum systems for high precision measurements, a whole research domain of quantum metrology has emerged.
Limitations to measurement precision from quantum mechanical uncertainties have been investigated and protocols to use measurements to optimally distinguish differently prepared quantum state have been developed, cf. \cite{braunstein_statistical_1994,wiseman_quantum_1996,Boixo2008,Grond2011,Lucke2011,Giovannetti2004}.

The continuous observation of a quantum system involves leakage of information via coupling of the system to a suitable meter, and an archetypal example is that of measurements of photons emitted from a quantum light source.
Laser spectroscopy thus involves the excitation of a quantum system, and detection of the fluorescence signal as function of laser frequency permits a fit, e.g., to a Lorentzian distribution and thus yields information about the resonance frequency and linewidth.
The resonance curve, however, represents only a part of the acquired data, as it omits details concerning the temporal dynamics and noise properties of the detection signal.
With the emergence of stochastic Schr\"odinger and master equations, which determine the quantum state conditioned on the full, noisy detection signals, it has been a natural next step to develop strategies to extract information from continuously probed systems.
Immediate applications then concern sensing of the magnitude of perturbations acting on the system, such as the magnetic field probed in atomic magnetometers \cite{Wasilewski2010,Shah2010}, and near field effects, e.g., from nuclear spins, probed by a single NV-center in diamond \cite{Kolkowitz2012,Zhao2012}.
Theoretical strategies have been proposed to continuously update parameter estimates based on the acquired data in a Bayesian manner on equal footing with the conditioned quantum state of the probe system, \cite{Gambetta2001,Negretti2012}.
The latter method is particularly useful, if the system can be approximated by Gaussian states \cite{Petersen2005,Geremia2003}.

The purpose of the present paper is to provide a formal link between some of the central ideas in classical estimation theory and stochastic master equations, and to identify efficient and systematic means to estimate unknown parameters from quantum measurement records.
We will, in particular, discuss and demonstrate methods applicable in cases where the parameter space is too large to permit a recursive Bayesian update procedure.
The methods are general, but for concreteness, we will consider light emitting quantum systems, and we will present explicit analyses and results for direct photon detection and for homodyne detection of the emitted radiation.

In photon counting, the measurement signal is a discrete process $N_t$, characterized by click events at specific times, and the density matrix $\rho_t$ of the emitter conditioned on the detection signal until time $t$ satisfies the non-linear filter equation
\begin{multline}
 d\rho_t = \left[ -i\cm{H, \rho_t} - \frac{1}{2}\acm{c^\dagger c, \rho_t} + \trace(c^\dagger c \rho_t)\rho_t \right] dt + \\
  \left[ \frac{c \rho_t c^\dagger}{\trace(c^\dagger c \rho_t) } - \rho_t \right] dN_t, \label{eq:JumpFilter}
\end{multline}
where the differential measurement result $dN_t$ is a Poisson increment, which is either $0$ (no click event) or $1$ (detector click event).
For the special case of a two level atom with upper (lower) states $\ket{e(g)}$ and with upper state lifetime $1/\gamma$, the conditional expectation $\Expectation[ dN_t | N_t ] = \trace(c^\dagger c \rho_t) dt$, where $c=\sqrt{\gamma}\ket{g}\bra{e}$.
The time-evolution of $\rho_t$ is therefore piecewise continuous (when $dN_t = 0$), but interrupted by jumps $\rho_t \mapsto c\rho_t c^\dagger / \trace(c^\dagger c\rho_t)$ at discrete times (when $dN_t = 1$).

It is also possible to perform field amplitude measurements by homodyne and heterodyne detection.
The measurement signal $Y_t$ is then a continuous function of time, and, e.g., when homodyne detection is performed on the fluorescence emitted by the decaying two-level atom, the conditioned density matrix satisfies an It\^{o} stochastic differential equation
\begin{multline}
 d\rho_t = \left[-i\cm{H, \rho} - \acm{c^\dagger c, \rho}/2 + c \rho c^\dagger \right] dt + \\
 (\M(\rho_t) - \trace(\M(\rho_t))\rho_t) (dY_t - \trace(\M(\rho_t)) dt), \label{eq:DiffusionFilter}
\end{multline}
where $\M(\rho) = c\rho + \rho c^\dagger$.
The differential measurement result $dY_t$ satisfies $dY_t = \trace(\M(\rho_t)) dt + dW_t$, where $dW_t$ is a Wiener increment with zero mean and variance $dt$.

In a generic experiment, all terms in the stochastic master equation can be parametrized by a vector of classical variables $\theta \in \mathbb{R}^n$, such as laser-atom detunings, which may in turn depend on the unknown values of externally applied fields, decay rates, temperature, etc.
In order to solve Eqs. (\ref{eq:JumpFilter}, \ref{eq:DiffusionFilter}), candidate values for these parameters need to be specified, and the goal of parameter estimation by continuous quantum measurements is to identify the best candidate values for the parameters $\theta$, given the actual measurement record ($N_t$ or $Y_t$).

In Sec. II, we review general parameter estimation concepts relevant to this manuscript: Bayesian inference, likelihood functions, and the Fisher information.
In Sec. III, we show how likelihood functions and the Fisher information can be efficiently obtained from the solution of the stochastic master equation of continuously monitored quantum systems.
In Sec. IV, we introduce the Markov chain Monte Carlo method for efficient sampling of the likelihood function in large search spaces, and we give numerical examples which illustrate the application and the results of our methods.
In Sec. V, we present a conclusion and outlook.

\section{Bayesian inference} \label{sec:BayesianInference}

Our theory of estimation is based on Bayes rule,
\begin{align}
 P(\theta| D) = \frac{P(D | \theta ) P(\theta)}{P(D)}, \label{eq:BayesRule}
\end{align}
where $P(\theta | D)$ is the probability density of the parameters $\theta$, given the observed data $D$.
Informally, this object contains all the information about the system parameters $\theta$ contained in the observed data $D$.
From this distribution we can calculate any estimate of interest, including the mean value, the mode and quantiles.
An important advantage of calculating the full probability density $P(\theta| D)$ is that it explicitly contains information about the uncertainty of the estimates.

Bayes rule relates the conditional probability density to the probability of observing the data given the parameters $P(D|\theta)$ and the existing prior information about the parameters $P(\theta)$.
The difficulty in using Eq. (\ref{eq:BayesRule}) stems from the denominator being a weighted integral over all possible parameter values $P(D) = \int d\theta P(D|\theta)P(\theta)$.
This integral is high-dimensional when several parameters are estimated, and the integrand can vary many orders of magnitude.

To determine $P(\theta|D)$ in practice, we therefore need a method for calculating $P(D|\theta)$ and a numerically efficient method of calculating $P(D)$.

\subsection{Likelihood functions} \label{sec:Likelihood}

Since the data $D$ is a measurement record, i.e., a function of time, its probability density, or \emph{likelihood}, $P(D|\theta)$ is difficult to define.
Apart from its use in the Bayesian update rule (\ref{eq:BayesRule}), it is common to maximize the likelihood with respect to the parameters $\theta$ and thus to estimate the true value of the parameter by the value for which the likelihood for generating the data is highest.

Instead of maximizing $P(D|\theta)$ with respect to $\theta$, one may maximize the value of any strictly increasing function $f$ of $P(D|\theta)$.
The logarithm is commonly used, and the resulting function is then denoted the \emph{log-likelihood} function.

It is also possible to divide $P(D|\theta)$ with any strictly positive function $P_0(D)$, without changing the location of the maximum with respect to $\theta$.
Thus any function $f( P(D|\theta) / P_0(D) )$ can be used to determine the maximum likelihood estimate of $\theta$.
We will use the term \emph{likelihood function} for any function $L(D|\theta) = P(D|\theta) / P_0(D)$ and logarithmic likelihood for $l(D|\theta) = \log L(D|\theta)$.
The likelihood function $L(D|\theta)$ can to a large extent be chosen to have a convenient form.

Since $P(D) = \int d\theta P(D|\theta) P(\theta) = P_0(D) \int d\theta L(D|\theta) P(\theta)$ we can rewrite Eq. (\ref{eq:BayesRule}) as
\begin{align}
 P(\theta|D) = \frac{ P(D|\theta) P(\theta)}{P_0(D) \int d\theta L(D|\theta) P(\theta) }=
 \frac{L(D|\theta) P(\theta)}{ \int d\theta L(D|\theta) P(\theta) }.
\end{align}

In the following we shall calculate the likelihood associated with continuously monitored light emitting quantum systems, where the function $P_0(D)$ is the probability density for either a Poisson- or a Wiener-process.

\subsection{Fisher information} \label{sec:FisherTheory}
A reasonable question to ask is, how accurate is the Bayesian estimate on average, and what is the fundamental limit on how accurate it is possible to estimate $\theta$?

The answer to this question is given by the Fisher Information matrix.
The Fisher Information matrix is defined in terms of the probability density for the data given some parameter $\theta$ and $P(D|\theta)$ as
\begin{align}
 I = \Expectation\left[ \left( \frac{\partial \log P(D|\theta)}{\partial\theta} \right)^2 \right], \label{eq:BasicFisher}
\end{align}
where the expectation is over all possible realizations of the data $D$.
The Cramér-Rao bound \cite{cramer_mathematical_1954} states, that any estimator for $\theta$, $\hat\theta(D)$ has a variance larger than $1/I(\theta_0)$, where $\theta_0$ is the true value of $\theta$.

If one uses a uniform prior, the Fisher Information of $P(D|\theta)$ is the reciprocal of the width of $P(\theta|D)$ averaged over the possible measurement records.
If a non-uniform prior is included, the reciprocal width of $P(\theta|D)$ is then, qualitatively, the sum of the Fisher Information for $\theta$ and the reciprocal width of the prior.

As described in section \ref{sec:Likelihood}, we can use any likelihood function in place of the conditional probability $P(D|\theta)$.
The same result holds for the Fisher information.
That is, in Eq. (\ref{eq:BasicFisher}) we can use any likelihood function $L(D|\theta)$ instead of the probability density.

For multiple variables, the Fisher information is
\begin{align}
 I_{ij} &= \Expectation\left[ \frac{\partial \log L(D|\theta) }{\partial\theta_i} \frac{\partial \log L(D|\theta) }{\partial\theta_j} \right], \label{eq:GeneralFisher} \\
  &=\Expectation\left[ L(D|\theta)^{-2} \frac{\partial L(D|\theta) }{\partial \theta_i} \frac{\partial L(D|\theta) }{\partial \theta_j} \right] \label{eq:GeneralFisher2}
\end{align}
where $L(D|\theta) $ is a likelihood function for observing $D$ given the parameters $\theta$.
The Cramér-Rao bound now states (for any unbiased estimator $\hat\theta$), that $\Expectation[ (\hat\theta_i - \theta_i^0) (\hat\theta_j - \theta_j^0) ] \geq ( I(\theta)^{-1} )_{ij} $.

\section{Continuously monitored quantum systems} \label{sec:Continously-monitored-systems}

The jump and diffusion quantum filter equations (\ref{eq:JumpFilter}, \ref{eq:DiffusionFilter}) are special cases of the general transformation of open quantum system density matrices subject to the random back action of measurements.
If, at a given instant of time, a measurement occurs with outcome $m \in M$, there is an effect-operator $\Omega(m)$ associated with each outcome, so that the state, conditioned on the outcome $m$ reads,
\begin{align}
 \rho|m = \frac{\Omega(m) \rho \Omega^\dagger(m)}{\trace(\Omega^\dagger(m) \Omega(m) \rho)}. \label{eq:FullConditional}
\end{align}
The probability (density) for observing the result $m$ is
\begin{align}
 p(m) = \trace(\Omega^\dagger(m) \Omega(m) \rho), \label{eq:usualp}
\end{align}
and the effect operators obey the relation
\begin{align}
 \int dm \Omega^\dagger(m)\Omega(m) = \id.
 \end{align}

The normalization factors in Eq. (\ref{eq:FullConditional}) are exactly the probabilities (\ref{eq:usualp}) to obtain the corresponding measurement outcome, and a similar probabilistic interpretation holds for the non-linear terms including the coefficients $\trace(c^\dagger c\rho_t)\rho_t$ and $\trace(\M(\rho_t))\rho_t$, in Eqs. (\ref{eq:JumpFilter}, \ref{eq:DiffusionFilter}).
This implies that if the stochastic density matrix equation is solved without incorporating the renormalization factors, the decreasing trace of $\rho$ with time yields the likelihood for the actual detection record to occur.

In the quantum jump master equation, the jump probability, and hence the decrease in density matrix norm associated with a single jump is proportional to the duration of the infinitesimal time step $dt$ chosen for the simulation.
This causes an undesired and inconvenient dependence of the likelihood function on $dt$ and on the number of click events, that we can, however, eliminate by a simple extension of the theory \cite{wiseman_quantum_1996} similar to \cite{Goetsch1994}.

We introduce an arbitrary positive function $p_0(m)$ and rescale the effect operators $\Omega(m)\rightarrow \Omega(m)/\sqrt{p_0(m)}$ so that they now obey the modified normalization condition,
\begin{align}
 \int_M dm p_0(m) \Omega^\dagger(m) \Omega(m) = \id,
\end{align}
Eq. (\ref{eq:FullConditional}) still holds, but the probability distribution for the different outcomes factors
\begin{align}
 p(m) = p_0(m) \trace(\Omega^\dagger(m) \Omega(m) \rho), \label{eq:RadonNikodym}
\end{align}
and we have the freedom to choose the un-normalized conditional states,
\begin{align}
 \tilde\rho|m = \Omega(m) \rho \Omega^\dagger(m),
\end{align}
whose trace depends now explicitly on the chosen function $p_0(m)$.

The expectation value, denoted by $\Expectation$, of any function $f(m)$ is given by
\begin{multline*}
 \Expectation[ f(m) ] = \int_M dm p(m) f(m) \\
  = \int_M dm p_0(m) \trace(\tilde\rho|m) f(m) \equiv \Expectation_0[ \trace(\tilde\rho|m) f(m) ],
\end{multline*}
where $\Expectation_0$ is to be understood as the expectation with respect to the reference probability $p_0$.
In the following, we will suppress the dependence on the measurement outcomes and simply write $\tilde\rho$ rather than $\tilde\rho|m$ for the conditioned density matrix.

The trace of the conditioned state is renormalized by a factor that depends on the specific detection record and which does not change its relative dependence on different values of the unknown parameters $\theta$.
It thus still serves as a likelihood function for the Bayesian determination of their values.
Our scaling with the function $p_0(m)$ in Eq. (\ref{eq:RadonNikodym}) is indeed equivalent to the scaling allowed in the definition of the likelihood function, $L(D|\theta) = P(D|\theta)/P_0(D)$, in Sec. \ref{sec:Likelihood}.
The ``ostensible probability'' $p_0$ \cite{wiseman_quantum_1996} provides a reference measure $p_0(m) dm$ on the set of measurement outcomes, and for our application it serves as a convenient unit for the effect operators $\Omega(m)$.
The relative entropy of $p$ with respect to $p_0$ is given directly in terms of $\trace(\tilde\rho)$ as $S(p|p_0) = \Expectation[\log(\trace(\tilde\rho))]$.

Describing continuous measurements as the $N\rightarrow \infty$ limit of a process of $N$ times repeated measurements, it is natural to consider a general reference probability distribution on $M^N$, such that the probability of the measurement record factors,
\begin{multline}
 p(m_1, \ldots m_N) = p^{(N)}_0(m_1, \ldots m_N) \\
\trace( \Omega(m_N) \ldots \Omega(m_1) \rho \Omega^\dagger(m_i) \ldots \Omega^\dagger(m_N) ),
\end{multline}
where convenient reference probabilities $p^{(N)}_0(\vec m) = p_0(m_1) \ldots p_0(m_N)$ for the jump-type and diffusion-like measurements will be given below.

\subsection{Jump type equation}

\begin{figure*}
  \includegraphics{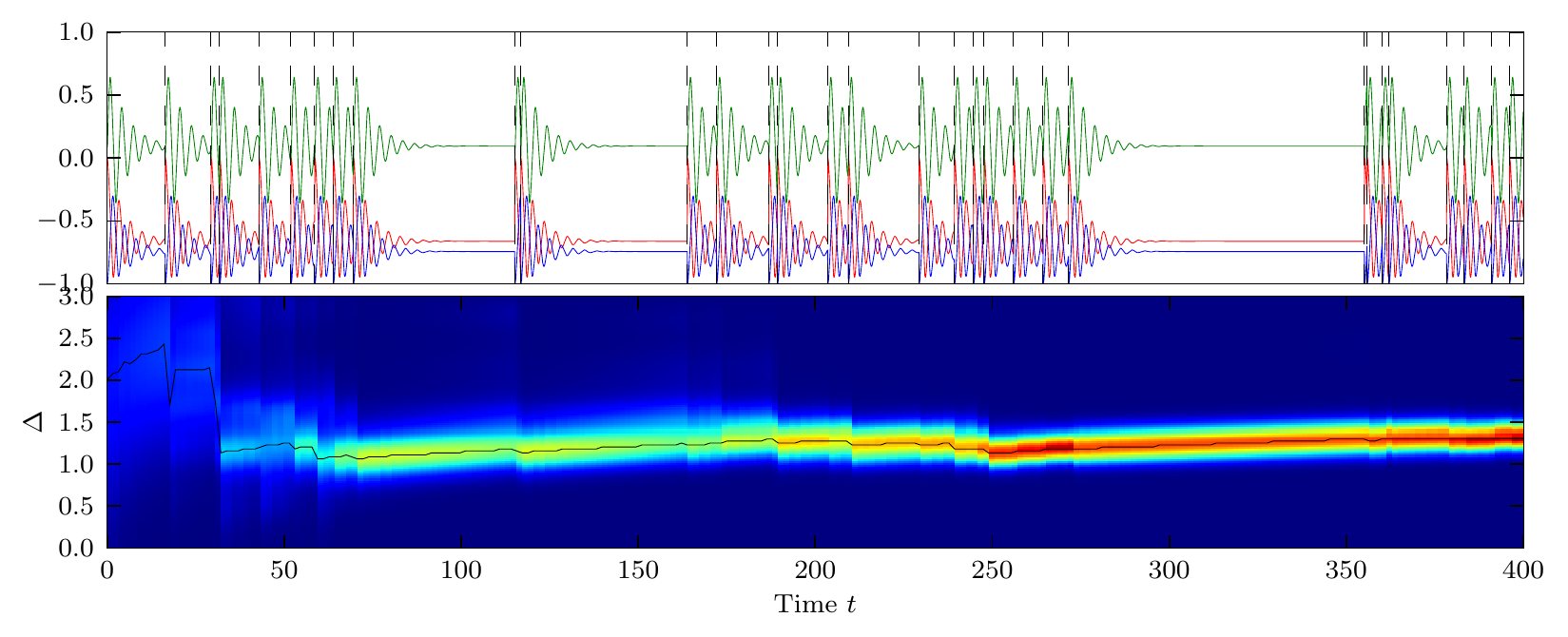}
\caption{(Color online) The upper panel shows the three components of the Bloch vector for a two-level atom subject to laser excitation with Rabi frequency $\Omega = 1.3$ and detuning $\Delta = 1.43$ (dimensionless units).
The atomic inversion is represented by the lower (blue) curve.
The atom decays with a rate $\gamma = 0.55$, leading to the observation of quantum jumps at the instants indicated by vertical dashed lines and the transient Bloch vector dynamics.
The lower panel shows the estimation of the detuning $\Delta$, treated as an unknown variable with a probability distribution, which is updated in Bayesian manner, conditioned on the measurement record.
See text.} \label{fig:RotspinExampleTrajectory}
\end{figure*}

For the jump-type measurements there are for each small time interval $dt$ two possible detector outcomes, $dN_t = 0$ and $dN_t = 1$.
We use our freedom to choose $p_0$ as the probability for a Poisson process with rate $\lambda$, i.e. $p_0(dN_t = 1) = \lambda dt$ and $p_0(dN_t = 0) = 1 - \lambda dt$, and the correspondingly normalized measurement effect operators
\begin{align}
\begin{split}
 \Omega_0 &= \id - i H dt - \frac{1}{2} (c^\dagger c - \lambda) dt \\
 \Omega_1 &= \frac{c}{\sqrt{\lambda}}
\end{split}
\end{align}
The probability for a detector click is $p_0(dN_t = 1)\trace( \Omega_1^\dagger \Omega_1 \rho_t) = \trace(c^\dagger c \rho_t) dt$ as expected from a rate process, and the expected number of events is $\Expectation[dN_t|N_t] = \trace(c^\dagger c \rho_t) dt$, while the reference expected value is $\Expectation_0[dN_t|N_t] = \lambda dt$.

The un-normalized conditional quantum state can be expressed as follows
\begin{multline}
 d\tilde\rho_t = \left[ -i\cm{H, \tilde\rho_t} - \frac{1}{2}\acm{c^\dagger c, \tilde\rho_t} + \lambda\tilde\rho_t \right] dt \\
  + dN_t \left[ \frac{c \tilde\rho_t c^\dagger}{\lambda} - \tilde\rho_t \right], \label{eq:JumpLinear}
\end{multline}
while explicit normalization leads to Eq. (\ref{eq:JumpFilter}).

The dynamics of $\tilde\rho_t$ is governed by the Hamiltonian $H$ and the operator $c$ which in turn depend on the parameters $\theta$.
The likelihood of a specific sequence of detection events at times $t_1, \ldots t_N < t$ is simply $L( t_1, \ldots t_N | \theta) = \trace(\tilde\rho_t)$, and Eq. (\ref{eq:JumpLinear}) thus provides a differential equation for the likelihood function $L_t = \trace(\tilde\rho_t)$
\begin{align}
 d L_t &= (\lambda L_t - \trace(c^\dagger c \tilde\rho_t)) dt + dN_t \left[ \frac{ \trace(c^\dagger c \tilde\rho_t) }{\lambda} - L_t \right], \label{eq:like-lin}
\end{align}
where we have suppressed $L_t$'s dependence on $\theta$.

The solutions $\rho_t$ of Eq. (\ref{eq:JumpFilter}) and $\tilde{\rho}_t$ of Eq. (\ref{eq:JumpLinear}) obey $\tilde{\rho}_t=\trace(\tilde\rho_t) \rho_t = L_t \rho_t$ which can be inserted in (\ref{eq:like-lin}) to yield,
\begin{align}
 d L_t &= (\lambda - \trace(c^\dagger c \rho_t)) L_t dt + dN_t \left[ \lambda^{-1} \trace(c^\dagger c \rho_t) - 1 \right] L_t. \label{eq:JumpNotLog}
\end{align}
This shows that even though the likelihood is formally defined by the trace of the un-normalized conditioned density matrix, $L_t$ can be calculated from the normalized state $\rho_t$ satisfying Eq. (\ref{eq:JumpFilter}).

For numerical purposes it is convenient to work with $l_t = \log L_t$ which satisfies
\begin{align}
 dl_t = (\lambda - \trace(c^\dagger c \rho_t)) dt + dN_t \log(\trace(c^\dagger c\rho_t)/\lambda). \label{eq:JumpLikelihood}
\end{align}

\subsection{Diffusion equation}
For diffusion type measurements, describing, e.g., homodyne detection of light, the set of outcomes in a small time interval $dt$ is the real numbers.
We will here use the probability of a Wiener increment $dW_t$, i.e. a normal distribution with mean zero and variance $dt$ as our reference probability $p_0^W$.
The effect of observing a result $dY_t$ is
\begin{align}
 \Omega(dY_t) = \id - i H dt - \frac{1}{2} c^\dagger c dt + c dY_t,
\end{align}
and the probability for observing a given value $dY_t$ is
\begin{align}
 p(dY_t) = p_0^W(dY_t) (1 + \trace((c + c^\dagger)\rho_t) dY_t).
\end{align}

We can calculate $\Expectation[dY_t | Y_t] = \Expectation_0[ (1 + \trace((c + c^\dagger)\rho_t) dY_t) dY_t | Y_t] = \trace((c + c^\dagger)\rho_t) dt$ and $\Expectation[dY_t^2|Y_t] = \Expectation_0[ (1 + \trace((c + c^\dagger)\rho_t) dY_t) dY_t^2 | Y_t] = dt$, which implies
\begin{align}
 dY_t = \trace((c + c^\dagger)\rho) dt + dW_t, \label{eq:DiffusionMeasurementResult}
\end{align}
where $dW_t$ is a Wiener increment with respect to the full probability distribution $p$ (while $dY_t$ is a Wiener increment with respect to $p_0$).

The un-normalized stochastic differential equation becomes
\begin{multline}
 d\tilde\rho_t = \left[ -i\cm{H, \tilde\rho_t} - \frac{1}{2}\acm{c^\dagger c, \tilde\rho_t} + c \tilde\rho_t c^\dagger \right] dt \\
  + (c \tilde\rho_t + \tilde\rho_t c^\dagger) dY_t, \label{eq:DiffusionLinear}
\end{multline}
and it leads to the likelihood $L_t = \trace(\tilde\rho_t)$ satisfying
\begin{align}
 dL_t = \trace( \M(\tilde\rho_t ) ) dY_t.
 \end{align}

As above, we can also express $L_t$ in terms of the normalized solution to Eq. (\ref{eq:DiffusionFilter}),
\begin{align}
 dL_t = \trace( \M(\rho_t) ) L_t dY_t, \label{eq:DiffusionNotLog}
\end{align}
and the log-likelihood $l_t = \log L_t$ satisfies
\begin{align}
 dl_t = \trace(\M(\rho_t)) (dY_t - \trace(\M(\rho_t)) dt). \label{eq:DiffusionLikelihood}
\end{align}

\subsection{Fisher information}

Using $\trace(\tilde\rho_t)$ as our likelihood function, we can apply Eq. (\ref{eq:GeneralFisher2}) to calculate the Cramér-Rao bound for estimating the unknown parameters in the system dynamics under both types of measurements.
Define the matrices
\begin{align}
 \rho^i_t = \frac{1}{\trace(\tilde\rho_t)} \partial_i \tilde\rho_t,
\end{align}
where the derivative is with respect to the $i$'th component of the vector of parameters $\theta$.
The expectation value of $\trace(\rho^i_t)\trace(\rho^j_t)$ with respect to the probability distribution $p$ (i.e. the actual probability for generating a trajectory) will then be the $ij$-component of the Fisher Information matrix for the continuously monitored quantum system.
We can therefore evaluate the Fisher information matrix numerically by simulating the stochastic master equation a large number of times and determine the expectation value Eq. (\ref{eq:GeneralFisher2}).

In practice, for the jump-type measurement, this requires solution of Eq. (\ref{eq:JumpFilter}) together with a simultaneous evaluation of the matrices $\rho^i_t$, which can, in turn, be determined from the inhomogeneous jump type master equation
\begin{multline}
 d\rho^i_t = \left[ -i\cm{H, \rho^i_t} - \frac{1}{2}\acm{c^\dagger c, \rho^i_t} + \trace(c^\dagger c \rho^i_t)\rho^i_t \right] dt \\ + \left[ -i\cm{\partial_i H, \rho_t} - \frac{1}{2}\acm{\partial_i(c^\dagger c), \rho^i_t} \right] dt \\
  + dN_t ( c\rho^i_t c^\dagger + (\partial_i c)\rho_t c^\dagger + c\rho_t (\partial_i c^\dagger) - \rho^i_t ), \label{eq:FisherEqJump}
\end{multline}
where the stochastic term $dN_t$ takes the same value as in Eq. (\ref{eq:JumpFilter}), and where the derivative of the Hamiltonian and damping terms with respect to $\theta_i$ are assumed known.

Similarly, for the diffusion-type measurement
\begin{multline}
 d\rho^i_t = \left[-i\cm{H, \rho^i_t} - \acm{c^\dagger c, \rho^i_t}/2 + c \rho^i_t c^\dagger \right] dt\\
 + \left[ -i\cm{ \partial_i H, \rho_t} -\acm{\partial_i(c^\dagger c), \rho_t}/2 + (\partial_i c)\rho_t c^\dagger +
 c \rho_t (\partial_i c^\dagger )\right] dt\\
  (\M(\rho^i_t) + (\partial_i\M)(\rho_t) - \trace(\M(\rho_t)) \rho^i_t) (dY_t - \trace(\M(\rho_t)) dt), \label{eq:FisherEqDiffusion}
\end{multline}
where the Wiener increment $dY_t - \trace(\M(\rho_t)) dt = dW_t$ takes the same value as in Eq. (\ref{eq:DiffusionFilter}).

The Fisher information provides an average quantifier of the asymptotic uncertainty in the estimation problem.
With Eqs. (\ref{eq:JumpFilter}, \ref{eq:FisherEqJump}) and Eqs. (\ref{eq:DiffusionFilter}, \ref{eq:FisherEqDiffusion}) we have shown how the Fisher information can be calculated by simulating many independent sequences of the stochastic master equation for the two different types of measurement.
These simulations have to be carried out for the candidate values of the parameters to yield the precision expected for an estimate based on a typical experimental run.
As illustrated by comparison of other such precision measures in \cite{Gambetta2001}, different measurement schemes have different resolving power, and in future work, we plan to address these differences in more detail, e.g., by comparing the Fisher information derived for the jump-type and for different diffusion-type measurements.

We also note, that if the field/meter degrees of freedom could be left unmeasured, the full entangled density matrix of the quantum system and the quantized radiation field would depend on the unknown parameters.
Thus the general quantum Cramér-Rao bound derived by Braunstein and Caves \cite{braunstein_statistical_1994} to determine a parameter, encoded in a quantum state, yields the ultimate accuracy with witch the parameters in our state dynamics can be inferred using any type of measurements.
Identifying that accuracy, and investigating how closely it is approached by quantum jump and quantum diffusion measurements of the emitted light presents an interesting challenge for further studies.

\section{Numerical investigation}

In this section we will illustrate the theory outlined in the previous sections with a few characteristic examples.
One approach for investigating $P(\theta|D)$ is to compute the likelihood function $L(D|\theta)$ on a grid.
Using such a calculation, posterior expectation values of $\theta$ can be calculated by numerical integration.
A numerical maximization routine can also be used to find the maximum of $L(D|\theta)$ and thus provide a maximum likelihood estimate of the parameters.
The uncertainty is given by the Fisher information found by solution of the stochastic master equation with samples of simulated detection records.
The posterior probability density may have many local maxima and it can be difficult to find the global maximum of $L(D|\theta)$ using standard maximization techniques.

If the parameter space is very large, more efficient methods for sampling the likelihood function exist.
To sample a function with an un-normalized probability density $\pi(x)$, one can apply a random process for the candidate values in the form of a Markov chain, where the values jump in an appropriately chosen manner so that they attain the correct relative probabilities.
The transition probability $t(x_1 \to x_2)$ must hence be chosen such that it asymptotically reproduces the relative probability density $\pi(x)$.
The requirement for the transition rule $t$ is then that the only function that satisfies $\int dx f(x) t(x \to x') = f(x')$ is proportional to our desired $\pi(x)$.
A generic way to construct such a Markov chain is the Metropolis-Hastings algorithm \cite{press2007numerical,Gilks1995} which is used in many areas of science, and we provide a brief review of our application of the method.

The basic idea is to compare the relative probability densities of a randomly chosen candidate value $x_2$ with the one of the current value $x_1$.
The value $x_2$ may be chosen randomly or, more conveniently, according to a \emph{proposal chain} $q(x_1 \to x_2)$, e.g., in the neighborhood of $x_1$.
A correct sampling of the probability density is obtained by accepting $x_2$ with the probability
\begin{align}
  \alpha(x_1, x_2) = \min\left(1, \frac{\pi(x_2)q(x_2 \to x_1)}{\pi(x_1) q(x_1 \to x_2)}\right), \label{eq:MCMCjumpP}
\end{align}
and otherwise retaining the value $x_1$.
If the proposal chain is able to explore the entire parameter space this Markov chain will have $\pi(x)$ as un-normalized stationary distribution.

A nice feature of the Metropolis-Hastings sampling method, is that it uses only ratios between different arguments of the functions $\pi$ and $q$.
This implies, that we can use the un-normalized probabilities $\pi(x)$, and for our purpose, we can use the likelihood functions found by solving (\ref{eq:JumpNotLog}) or (\ref{eq:DiffusionNotLog}) with the parameter values $\theta= x_1$ and $\theta=x_2$).

In summary, to sample the posterior density for the estimated parameters $P(\theta|D)$ for a continuous quantum measurement using Metropolis-Hastings we select a random $\theta$ from the prior distribution $P(\theta)$, and we proceed as follows:
\begin{enumerate}
 \item Determine candidate $\theta_c$ according to some proposal distribution $q(\theta \to \theta_c)$.
 \item Calculate the likelihood or, equivalently, the log-likelihood $l_T^c$ for the data until the final time $T$, using the candidate $\theta_c$ and Eqs. (\ref{eq:JumpNotLog}, \ref{eq:DiffusionNotLog}) or Eqs. (\ref{eq:JumpLikelihood}, \ref{eq:DiffusionLikelihood}) depending on the type of measurement.
 \item Calculate $\alpha(\theta, \theta_c) = \min(1, \exp(l_T^c - l_T) q(\theta_c \to \theta)/q(\theta \to \theta_c)$, where $l_T$ is the log-likelihood for the previous parameter $\theta$.
 \item Accept candidate with probability $\alpha(\theta, \theta_c)$, otherwise keep $\theta$.
\end{enumerate}
These steps are repeated a large number of times, and the parameters sampled are then representative and can be used for determination of any property of the distribution $P(\theta|D)$.

In the simulations presented below, the proposal distribution $q(\theta \to \theta_c)$ was chosen as a multivariate normal distribution centered at $\theta$ with a variance selected to achieve a reasonable acceptance rate of $10\%$ to $50\%$.

Many techniques exist for investigating the convergence rate and the correlation length of the Markov chain generated by the above technique \cite{Gilks1995}.
In the simple examples studied in the present manuscript, the convergence rate and correlation length are readily identified, but a more careful analysis of these issues is necessary when applying the technique to an experimental situation with many parameters and uncertainties.

\section{Examples}

\subsection{Two-level atom}

\begin{figure}
  \includegraphics{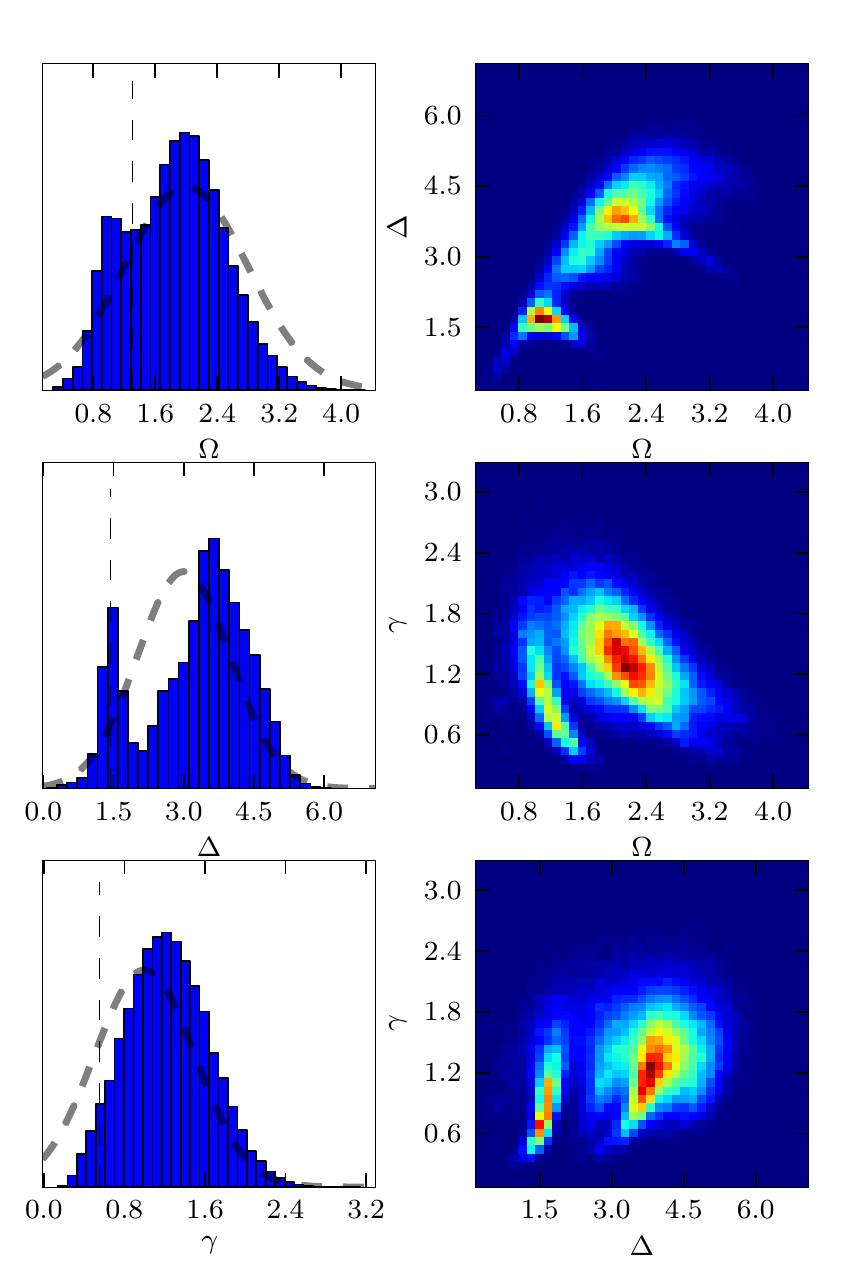}
\caption{(Color online) The left panels show histograms of Markov Chain Monte Carlo sampled distributions of the parameters, $\Omega,\ \Delta,\ \gamma$ in our two-level atom model.
The prior knowledge of the parameters assumes normal distributions, shown by the dashed lines, with mean values $\mu_\Omega = 2.0$, $\mu_\Delta = 3.0$, $\mu_\kappa = 1.0$ and standard deviations $\sigma_\Omega = 0.8$, $\sigma_\Delta = 1.0$ and $\sigma_\kappa = 0.5$.
The right panels display the correlations between the different pairs of sampled parameters.
}  \label{fig:TwoLevelMCMC}
\end{figure}

Consider a coherently driven two-level atom that decays by spontaneous emission of photons.
The atom is described by the Hamiltonian $H = (\Omega/2) \sigma^x + (\Delta/2) \sigma^z$ and by a jump operator $c = \sqrt{\gamma} \sigma^-$, where $\gamma$ is the effective decay rate, $\vec\sigma = (\sigma^x, \sigma^y, \sigma^z)$ is the vector of Pauli spin-matrices, and $\sigma^-$ denotes the Pauli lowering operator.
The measurement record is the times at which photons are detected with a photodetector.

The top part of Figure \ref{fig:RotspinExampleTrajectory} shows an example trajectory, assuming known values $\gamma = 0.55$, $\Omega = 1.3$ and $\Delta = 1.43$ for the atomic and field parameters (in dimensionless units, e.g., relative to the decay rate of another excited state in the same atom).
The continuous curves show the components of the Bloch vector $\vec{r}=\trace(\rho\vec\sigma)$, and they display continuous evolution disrupted at discrete times, where discontinuous quantum jumps of the state occur associated with the detector clicks.
In the bottom part of Figure \ref{fig:RotspinExampleTrajectory}, we have assumed that $\gamma$ and $\Omega$ are known, and we evaluate the probability distribution for the detuning parameter on a grid, assuming a prior normal distribution for $\Delta$ with a standard deviation $\sigma_\Delta = 1.0$ and mean value $\mu_\Delta = 2.0$.

The $\Delta$-distribution is conditoned on the same detection record as applied in the upper part of the Figure, and we observe how the no-click
periods cause a continuous change of the posterior density, while the discrete jumps are accompanied by more abrupt changes, until the distribution is well converged.
The importance of the use of the whole signal and not only the mean photodetection rate, is easily understood by the observation that following each quantum jump, the atomic density matrix describes a transient damped Rabi oscillation, and the temporal probability distribution for the subsequent jump event is periodically modulated.
Since the period of the transient modulation depends explicitly on $\Omega$ and $\Delta$ the actual occurrence of the next jump strongly favors (disfavors) certain values of $\Delta$ and causes the conditional increase (decrease) in the probability density at those values.

With a single unknown parameter, it is possible to compute the likelihood function on a fine grid, but if we pass to the larger parameter space of more unknown variables, we have recourse to more advanced search methods.
In Fig. \ref{fig:TwoLevelMCMC}, we show the results of running the Markov chain Monte Carlo-algorithm on the trajectory in Fig. \ref{fig:RotspinExampleTrajectory} with all three parameters $\Omega,\ \Delta,\ \gamma$ treated as unknown.
We assume normal distributed priors, shown with the dashed lines in the left panels of Fig. \ref{fig:TwoLevelMCMC}, and the histograms show the values for the three parameters sampled by the Markov chain.
Since the trajectory is quite short, there is not sufficient information to perfectly infer the values of the parameters, and the joint densities of pairs of variables in the right panels indicate that two islands of likely values of the set of parameters are not resolved by the measurements.

We have compared the distribution of time differences between click event in the rather short detection record, shown in Fig. \ref{fig:RotspinExampleTrajectory}, with the expected transient Rabi oscillation dynamics and we find that they are, indeed, compatible with the different values for the pair of parameters $\Omega,\ \Delta$, occurring with comparable probabilities in the upper right panel in Fig. \ref{fig:TwoLevelMCMC}.
With a few more "lucky clicks", however, the distribution will favor one choice, and due to the correlations between our estimates for all three parameters, they may then rather rapidly all converge to the correct values.

\begin{figure}
 \includegraphics[width=\columnwidth]{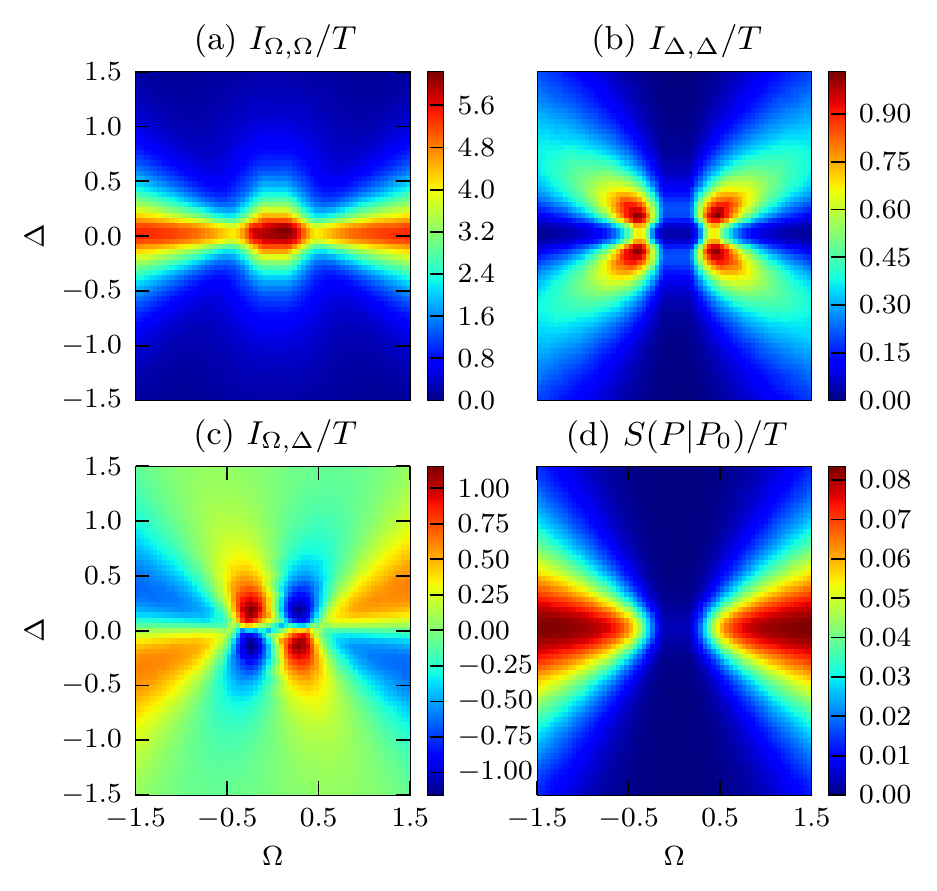}
 \caption{(Color online) Fisher Information matrix components for photodetection of a decaying two-level atom with decay rate $\gamma = 0.55$ initially in the unexcited state up to $T = 40$.
The two upper panels show the diagonal elements $I_{\Omega,\Omega}$ and $I_{\Delta,\Delta}$ and
the lower left panel shows the off-diagonal element $I_{\Omega,\Delta}$ of the Fisher information matrix.
The lower right panel, shows the relative entropy between the signal probability distribution $p$ and $p_0$,
where $p_0$ is a Poisson process of with the intensity of the stationary emission rate for the two-level atom $\lambda_{st} = \Omega^2 \gamma / (\gamma^2 + 4\Delta^2 + 2 \Omega^2 )$, and $\gamma = 0.55$, see text.
} \label{fig:SpinHalfFisher}
\end{figure}

We have also calculated the Fisher information matrix for a photon counting experiment by applying the simulation methods described above, and we obtain the results shown in Fig. \ref{fig:SpinHalfFisher}.
The Fisher information matrix was calculated by simulating the stochastic master equation and the associated equations for the $\rho^i_t$ for different choices of the parameters $(\Omega, \Delta) \in [-3/2, 3/2] \times [-3/2, 3/2]$, while the decay rate was assumed to be known and equal to $\gamma = 0.55$ in our dimensionless units and the initial atomic state was unexcited.

We recall, that the Fisher information, evaluated at the estimated values $\Omega$, $\Delta$ gives the uncertainties of these two quantities as well as their covariance.
It is not surprising that the sensitivity of the photon detection method depends on the actual values of the parameters.
The fact that $\Omega$ and $\Delta$ enter the problem as coefficients on non-commuting spin components in the atomic Hamiltonian suggests that spin uncertainty relations may result in limitations on their joint determination, see also \cite{Petersen2005}.
Such a fundamental limitation may be reflected by the apparent anti-correlation of the occurrence of large and small values of the Fisher information matrix elements $I_{\Omega,\Omega}$ and $I_{\Delta,\Delta}$ in Fig. \ref{fig:SpinHalfFisher}.

The relative entropy between the signal probability $p$ and a Poisson reference distribution $p_0$, $S(p|p_0) = \Expectation[\log L_t]$, i.e., the $p$-expectation value of $\log L_t$, is shown in Fig. \ref{fig:SpinHalfFisher}(d).
The reference distribution $p_0$ has been chosen as a Poisson process with a rate set by the stationary emission rate $\lambda_{st} = \Omega^2 \gamma / (\gamma^2 + 4\Delta^2 + 2 \Omega^2 )$.
The relative entropy is close to zero in large regions of the $\Omega$, $\Delta$ parameter space, indicating that the emission process is not very different from a Poisson process.
In the regions with $|\Omega|\geq 0.5$, $\Delta \approx 0$, the dynamics deviate significantly from a Poisson process due to the Rabi oscillations in the photon waiting-time distribution, and the ensemble of trajectories have a higher entropy.

\subsection{Bi-modal two-level atom}
\begin{figure}
 \includegraphics{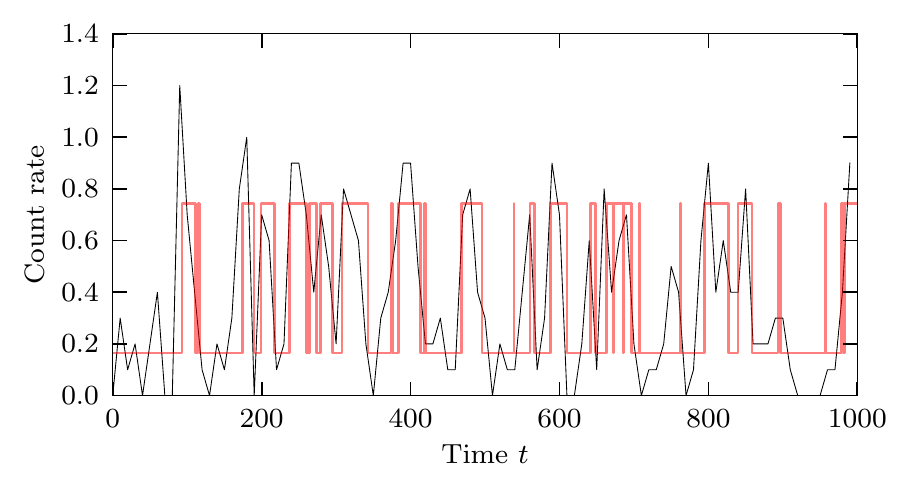}
 \caption{ (Color online) Simulated signal from a bimodal two-level atom, undergoing jumps and coherent evolution with two alternating sets of parameters, $a$ and $b$.
The solid black curve is the (binned) observed signal while the red dashed curve shows the mean expected counts for the atom subject to the current set of parameters.The values used for this trajectory are $\Omega_a = 1.1$, $\Delta_a = 1.3$, $\gamma_a = 1.6$, $\Omega_b = 2.2$, $\Delta_b = 0.2$, $\gamma_b = 2.4$ and transition rates $W(a \to b) = 0.03$ and $W(b \to a) = 0.08$.
 } \label{fig:TwoLevel-DoubleWellSignal}
\end{figure}

\begin{figure}
 \includegraphics{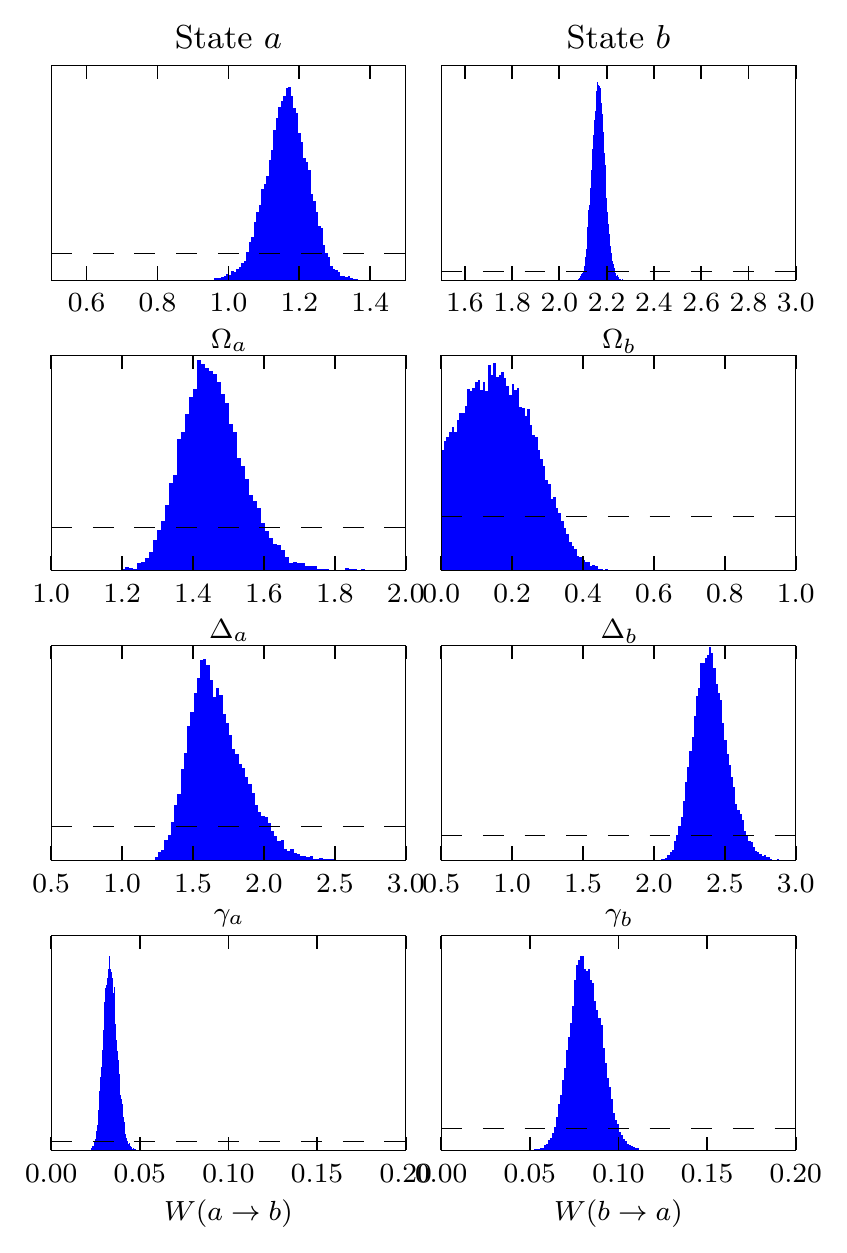}
 \caption{ (Color online) Marginal distributions for the eight unknown parameters in our bimodal two-level atomic system.
All prior distributions were taken to be uniform on the shown intervals as indicated by the black dashed line.
The estimation was based on the actual click events of the trajectory, partly shown in Figure \ref{fig:TwoLevel-DoubleWellSignal}.
 } \label{fig:TwoLevel-DoubleWellMCMC}
\end{figure}

Imagine now a situation, where the two-level atom is not subject to dynamics with a fixed set of unknown parameters, but it may jump randomly between two fixed sets of values.
Such jumps may occur due to changes in a binary variable in the surrounding environment. e.g., the quantum states $|a\rangle$ and $|b\rangle$ of a nearby atom, spin or mesoscopic qubit degree of freedom, or due to the atom moving spatially between two different positions in a laser field configuration.
We will assume these state changes are purely classical, i.e. we neglect all coherences between the configurations or positions $|a\rangle$ and $|b\rangle$, and we assume that both the Rabi-frequency, the detuning and the decay rate of the two-level atom have different values for the two states.

We describe the system using a conditional master equation where we include the environmental states $|a\rangle$ and $|b\rangle$ of the atoms in a block-diagonal density matrix, $\rho = \rho_a \tensorp \ket a \bra a + \rho_b \tensorp \ket b\bra b$, where $\rho_a$ ($\rho_b$) is the density matrix for the atom associated with the environmental state $a$ ($b$).
The system Hamiltonian is $H = \left( (\Omega_a/2) \sigma^x + (\Delta_a/2) \sigma^z\right) \tensorp \ket a \bra a + \left( (\Omega_b/2) \sigma^x + (\Delta_b/2) \sigma^z\right) \tensorp \ket b \bra b$ and the effective photo-detection jump operator is
\begin{align*}
 c = \sigma^- \tensorp (\sqrt{\gamma_a} \ket a \bra a + \sqrt{\gamma_b} \ket b \bra b).
\end{align*}
The transitions between the two configurations are described by incoherent jumping rates $W(a \to b)$ and $W(b \to a)$ and corresponding jump operators $J_{a\to b} = \sqrt{W(a\to b)} \id \tensorp \ket b \bra a$ and $J_{b\to a} = \sqrt{W(b\to a)} \id \tensorp \ket a \bra b$.
The system is now equivalent to an enlarged quantum system, and it is fully described as a single quantum system by the formalism outlined above.

We have used the parameters $\Omega_a = 1.1$, $\Delta_a = 1.3$, $\gamma_a = 1.6$, $\Omega_b = 2.2$, $\Delta_b = 0.2$, $\gamma_b = 2.4$ and (slow) transition rates $W(a \to b) = 0.03$ and $W(b \to a) = 0.08$ to simulate a typical detection record for the system.
In Figure \ref{fig:TwoLevel-DoubleWellSignal}, the black solid line shows the time-binned observed signal for this record as a function of time.
As the changes between the two sets of parameter occur at low rates, the photon counting permits efficient Bayesian determination of the classical states $a$ and $b$ along the same lines as \cite{Reick2010}.
The red curve shows the mean photon scattering rates evaluated for the current estimate of which set of parameters applies.

Treating all rates and coupling strengths as unknown, the large number of unknown parameters makes a straightforward Bayesian estimation of their values very complicated.
In Fig. \ref{fig:TwoLevel-DoubleWellMCMC} we show instead the outcome of the Markov Chain Monte Carlo sampling of the eight possible parameters over the same measurement sequence as in Fig. \ref{fig:TwoLevel-DoubleWellSignal}.
All values were assigned uniform prior probability distributions on the intervals shown (dashed lines in the figures), and the histograms show the concentration of the values sampled on the actual, correct parameters.

\section{Discussion and outlook}
In this paper we have presented a general method for inferring the values of parameters that govern the time-evolution of continuously monitored quantum systems.
The systems are described by stochastic master equations, and we have shown that the trace of the un-normalized density matrix can be interpreted and applied as a likelihood function in standard statistical methods for parameter inference.
Explicit differential equations for the likelihood function in terms of the normalized density matrix are exemplified for the case of photon counting (\ref{eq:JumpLikelihood}) and homodyne photodetection (\ref{eq:DiffusionLikelihood}).
The differential equations for the likelihood allows us to use numerically stable and efficient stochastic master equation simulations as input to a variety of standard statistical estimation algorithms, e.g. Markov Chain Monte Carlo and direct maximum likelihood estimation.
Our identification of the conditioned density matrix dynamics with the likelihood function, in addition, leads to an efficient method (\ref{eq:FisherEqJump}), (\ref{eq:FisherEqDiffusion}) to simulate the Fisher information associated with any particular measurement scheme, and thus to evaluate the confidence of parameter estimation by continuous measurements.

We presented our formalism for the case of photodetection, and in our examples we assumed that all emitted radiation is detected.
If there are unobserved decoherence or loss processes and, e.g., loss of the radiation signal before the detection, averaging over these processes simply contributes further (deterministic) dissipation terms of the Lindblad form $\mathcal{D}[J](\rho) = -\acm{J^\dagger J, \rho}/2 + J \rho J^\dagger$ in the master equations (1,2).
The likelihood equations (\ref{eq:JumpLikelihood},\ref{eq:DiffusionLikelihood}), however, remain unchanged.

A technical element in our formulation of the theory involves the introduction of a reference probability $p_0$, imposing a degree of freedom in the normalization of the effect operators and the density matrix conditioned on the measurement signal.
The introduction of the reference probability $p_0$ is mathematically equivalent to converting the set of measurement outcomes into a classical probability space with a reference probability measure $P_0$ and the quantum measurement effect operators induce a probability measure on $M$ via the relation $P(A) = \int_A dP_0(m) \trace(\tilde\rho|m)$ for subsets $A \subset M$.
In mathematical terms the likelihood function $\trace(\tilde\rho|m)$, discussed in section \ref{sec:BayesianInference}, can then be identified as the Radon-Nikodym derivative $dP/dP_0 (m)$.
This points to a further generalization by transforming the probability measure $P$ to some other measure $\tilde P$ such that $dP/d\tilde P = Z_t$, where $Z_t$ is a martingale, and where trajectories generated with the transformed probability measure $\tilde P$ should be weighted by $Z_t$ to obtain ensemble averages.
This may provide a useful technique to control the variance in numerically calculated ensemble averages and to simulate the master equation and the Fisher information matrix more efficiently.

The relative entropy $S(P|P_0)$ between the probability measures $P$ and $P_0$ is the $P$-expectation value of $\log L_t$, $S(P|P_0) = \Expectation[\log L_t] = \int dP \log(dP/dP_0)$ and we note that the Fisher information is nothing but the relative entropy between $P_\theta$ and $P_\theta'$ for infinitesimally close $\theta$ and $\theta'$.
Apart from their importance in parameter estimation, emphasized in this manuscript the entropy $S(P|P_0)$ and the Fisher information $I_{ij}$ provide means to characterize the stochastic dynamics of quantum trajectories in a manner similar to the use of entanglement susceptibility to characterize the correlations in quantum many-body physics \cite{zanardi_mixed-state_2007,gu_fidelity_2008,you_fidelity_2007,zanardi_information-theoretic_2007}.

\bibliography{biblio.bib}

\end{document}